\documentclass[a4paper,10pt,twocolumn]{article}

\usepackage{times}

\usepackage[latin1]{inputenc}
\usepackage{subfigure}
\usepackage{float}
\usepackage{graphicx}
\usepackage{amssymb}
\usepackage{amsmath}
\usepackage{booktabs}
\usepackage{arydshln}
\usepackage{url}

\newtheorem{query}{Hyp}
\newtheorem{stat}{Statistic}

\def\sw{{\sf sw}}
\def\rp{{\sf rp}}
\def\cp{{\sf cp}}

\def\<#1>{\langle#1\rangle}
\newtheorem{theorem}{Theorem}
\newtheorem{proposition}{Proposition}
\newtheorem{corollary}{Corollary}
\newtheorem{problem}{Problem}

\newcommand{\joinspace}{\negthickspace}
\newcommand{\jointwo}{\joinspace\Join\joinspace}
\newcommand{\joinone}{\Join\joinspace}

\usepackage{hyperref}
\usepackage{breakurl}

\usepackage{algpseudocode}
\usepackage[ruled]{algorithm}

\algnewcommand\algorithmicinput{\textbf{Input:}}
\algnewcommand\algorithmicoutput{\textbf{Output:}}
\algnewcommand\Input{\item[\algorithmicinput]}
\algnewcommand\Output{\item[\algorithmicoutput]}

\begin{document}

\title{Query Significance in Databases via Randomizations}

\author{ 
Markus Ojala \footnote{HIIT, Department of Information and Computer Science, Helsinki University of Technology, Finland} 
\and
Gemma C. Garriga \footnotemark[1] 
\and
Aristides Gionis \footnote{Yahoo! Research, Barcelona, Spain} 
\and
Heikki Mannila \footnotemark[1] 
}

\date{June 30, 2009}

\maketitle

\begin{abstract}

Many sorts of structured data are commonly stored in a multi-relational 
format of interrelated tables. 
Under this relational model, exploratory data analysis can be done
by using relational queries. As an example, in the Internet 
Movie Database (IMDb) a query can be used to check whether the average rank of 
action movies is higher than the average rank of drama movies.

We consider the problem of assessing whether the results returned 
by such a query are statistically significant or just a random artifact of the structure in the data. Our approach is based on randomizing the tables occurring 
in the queries and repeating the original query on the randomized tables. It turns out 
that there is no unique way of randomizing in multi-relational data. 
We propose several randomization techniques, study their properties, and 
show how to find out which queries or hypotheses
about our data result in statistically significant information. 
We give results on real and generated data and show how the significance of 
some queries vary between different randomizations.

\end{abstract}

\section{Introduction}

The question of evaluating whether certain hypotheses made from observed data are significant
or not, is one of the oldest problems in statistics. 
Statistical significance reduces an observed result (statistic) to a $p$-value that tells about
the probability of observing the same result at random when a certain null hypothesis is true. 
If this $p$-value is sufficiently small, we can assume that the null hypothesis is false. 
The technical challenge of defining an exact $p$-value for a given hypothesis is typically
resolved by studying the null distribution of the test analytically; for example,
the well known chi-squared test is based on statistics that follow a chi-square distribution
under the null hypothesis. Alternatively, when analytical solutions are not possible or
hard to state exactly, the null distribution can be defined via permutation tests. 

These useful statistical concepts have been used for years in experimental fields such as medicine,
biology, geology or physics, to name a few.
Many of these considerations have been extended as well to the data mining and database 
community. In a very first paper about association rules, Brin et al.~\cite{silverstein98beyond} 
considered measuring the significance of rules via the chi-squared test, and from there many
other papers followed---see e.g.~\cite{775053} for a comprehensive survey. 
More recently, the approach of defining randomization 
tests to assess data mining results was introduced for binary data~\cite{swaprand-gionisetal}, and for real-valued data~\cite{conf/sdm/OjalaVKHM08}.

\begin{figure}[t]
{\small
\begin{eqnarray*}
\mathrm{GM} & = & \{(\mathrm{Romance},m_1), (\mathrm{Romance},m_2),(\mathrm{Drama},m_3), \\ 
& & \ (\mathrm{Drama},m_4),(\mathrm{Drama},m_5),(\mathrm{Drama},m_6),  \\ 
& & \ (\mathrm{Drama},m_7),(\mathrm{History},m_6),(\mathrm{History},m_7)\} \\[.5em]
\mathrm{MD} & = &\{(m_1,\mathrm{C.~Waitt}), (m_2,\mathrm{C.~Waitt}), (m_3,\mathrm{C.~Waitt}), \\
& & \ (m_4,\mathrm{C.~Waitt}), (m_5,\mathrm{C.~Waitt}), (m_6,\mathrm{T.~George}), \\ 
& & \ (m_7,\mathrm{T.~George})\} \\[.5em]
\mathrm{DA} & = & \{(\mathrm{C.~Waitt},30), (\mathrm{T.~George}, 60)\} 
\end{eqnarray*}}
\vspace{-2em}
\caption{A toy example of a multi-relational database consisting of three binary relations: movies classified by genre, $\mathrm{GM}$; directors of movies, $\mathrm{MD}$; 
and ages of directors, $\mathrm{DA}$.}
\label{fig:toyex0}
\end{figure}

Abstracting a bit from the question of how significant patterns are in the data, we
introduce here 
the statistical testing framework to databases and the exploratory task of 
querying the relations of the database. The question of understanding what we know and what we believe 
about our dataset becomes 
tricky when the data is highly structured and interrelated.
Structured data is everywhere: examples are the Internet Movie Database (IMDb), or
the DBLP computer science bibliography, and indeed, most of today's information systems
are actually relational databases. In IMDb, 
e.g., 
basic entities
are directors, movies, genres, ranks or years; in addition, we have relations such as directors
direct movies, movies are classified 
by a genre, movies are ranked with some quality 
criteria, and directors are born in a certain year. Each of these relations is
represented in a separate table which relates to others through their common attribute values. 
A simple toy example is given in Figure~\ref{fig:toyex0}.

In multi-relational databases, users and applications access the data via queries. 
E.g., a query can be made to check the average age of directors of history movies, or the average 
age of directors of romance movies. In the toy example
of Figure~\ref{fig:toyex0}, the first query returns a value of 60,
while the second query returns a value of 30.  
Usually, the answer returned by the query is assumed as a fact, thus implying
some conventional wisdom---for this toy example we might be tempted to believe 
that directors of romance movies are younger than directors of history movies. 
But, should we really believe that this hypothesis is significant from the data? If we knew
that all history movies are also classified as drama movies, would the value of 60
still have the same importance? Or, if we knew that the same director has participated
in both romance and drama movies? 

We study whether the results returned by queries
are significant or just a random artifact due to the structure in the data. 
Our statistical tool is randomizations and the approach is simple: 
randomize certain relations occurring in the queries and
repeat the original query in the random samples. This provides an empirical $p$-value,
and, as in basic statistics, we can reject or accept our hypothesis linked to the query. 
The goal behind this idea is to provide an understanding of how the structure
of the data affects the significance of the information we derive from our queries. If certain
structures or patterns remain after simple randomizations (e.g.,\ the fact that history movies
are also drama movies in the toy example), the answers of a query that
rely on such patterns should be regarded as not significant.   

It turns out that there is no unique way 
of randomizing in multi-relational data, and indeed, it is difficult to give a 
fully satisfactory answer about which randomizations are more important than others. 
We study several randomization methods and show the combinatorial properties
of the null distributions on multiple tables. 
Our contribution makes a first step towards understanding how 
the significance of a query is linked to the structure hidden in the data;
randomizations are a sound statistical tool to make such a connection.  
We believe this is an important problem of interest to both the database and
data mining communities.
We present experimental results on synthetic data, and show
the usability of the method for several queries in real datasets.

\section{Problem statement}
\label{sect:probdef}

Let $A$ be a binary relation $A \subseteq I \times J$
between sets $I$ and $J$. 
In the market basket application, for example, $I$ could be 
a set of customers and $J$ a set of products. A binary relation 
$A \subseteq I \times J$ identifies which customers from $I$ buy which 
products from $J$. 
Notice that every binary relation can be 
seen as a binary matrix describing 
the occurrences between the row set $I$ and column set $J$, 
see Figure~\ref{fig:toyex1} for examples. 

Let $\{A_1,\ldots,A_n\}$ be a set of $n$ binary relations representing some structured data. 
This relational model is very general. It applies, for example, to a 
movie database system, as shown in Figure~\ref{fig:toyex1}. 
The representation of the same example 
as a sequence of bipartite graphs is depicted in Figure~\ref{fig:toyex1:graph}.

\begin{figure}[t]
\centering
\subfigure[Genre $\times$ Movie]{
\begin{tabular}{l|ccccccc}
\hline
& $m_1$ & $m_2$ & $m_3$ & $m_4$ & $m_5$ & $m_6$ & $m_7$ \\
\hline
{\small Romance} & 1 & 1 & 0 & 0 & 0 & 0 & 0 \\ 
{\small Drama} & 0 & 0 & 1 & 1 & 1 & 1 & 1 \\ 
{\small History} & 0 & 0 & 0 & 0 & 0 & 1 & 1 \\ 
\hline
\end{tabular}
}
\subfigure[Movie $\times$ Director]{
\begin{tabular}{c|cc}
\hline
& {\small C.~Waitt} & {\small T.~George}  \\
\hline
$m_1$ & 1 & 0 \\ 
$m_2$ & 1 & 0 \\ 
$m_3$ & 1 & 0 \\ 
$m_4$ & 1 & 0 \\ 
$m_5$ & 1 & 0 \\ 
$m_6$ & 0 & 1 \\ 
$m_7$ & 0 & 1 \\ 
\hline
\end{tabular}
}
\subfigure[Director $\times$ Age]{
\begin{tabular}{l|cc}
\hline
& $30$ & $60$ \\
\hline
{\small C.~Waitt} & 1 & 0 \\ 
{\small T.~George}  & 0 & 1 \\ 
\hline
\end{tabular}
}
\caption{The binary table representation of the toy database in Figure~\ref{fig:toyex0}: 
(a) GM; (b) MD; and (c) DA.}
\label{fig:toyex1}
\end{figure}

\begin{figure}[t]
\centering
\includegraphics[width=6cm]{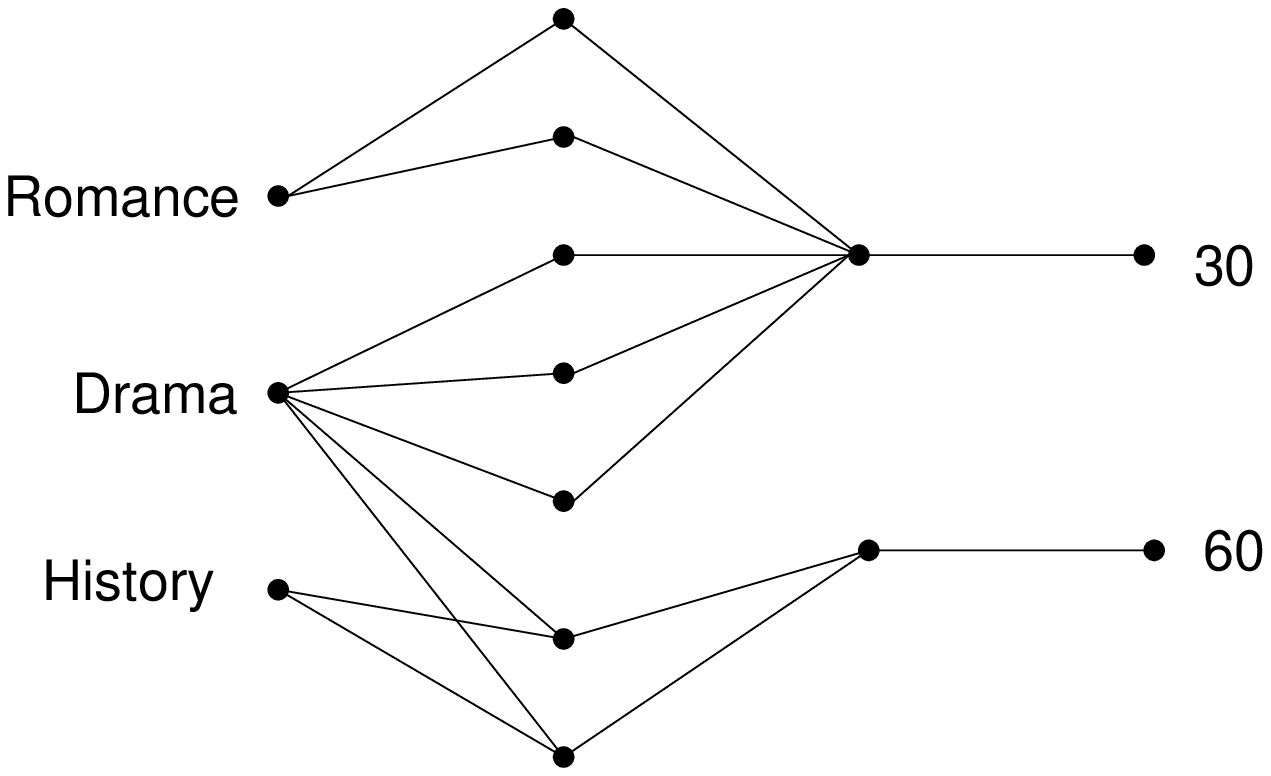}
\caption{The bipartite graph representation of the movie database shown in Figure~\ref{fig:toyex1}.
The graph shows all the possible paths from the source nodes, $\mathrm{Genre}$, to the destination nodes, $\mathrm{Age}$.}
\label{fig:toyex1:graph}
\end{figure}

The basic operator to combine relations is the {\it natural join}.
Conceptually, a join between two relations $A$ and $B$, denoted
$A \jointwo B$, combines all entries from $A$ and $B$ that share
common attribute values to return a composition 
of the relations. For example, given $(i,j) \in A$, $(j,k) \in B$ and $(j,k') \in B$,
we have $(i,j,k) \in A \jointwo B$ and also $(i,j,k') \in A \jointwo B$. 
The join operator is associative
over a set of relations and its result 
explicitly represents all existing paths between the occurring relations. 
For example, the natural join of the three tables in 
Figure~\ref{fig:toyex1} returns a tuple for each path 
there is between Genre and Age.
For an {\it ordered} subset of binary relations from the database
$S \subseteq \{A_1,\ldots,A_n\}$, we use $\joinone S$ to denote the
final join between all elements in $S$. The order in $S$ is to ensure
a join of consistent relations; we assume that $S$ in $\joinone S$
is always implicitly ordered.

A {\it query} $q$ is applied to the join of a subset of the relations in the database
$S \subseteq \{A_1,\ldots,A_n\}$. The result of a query is denoted by $q(\joinone S)$.
We say that $S$ is the set of relations occurring in the query.  
A query can be described with the operators of
projection and selection~\cite{ramakrirshnan03}, applied to a join $\joinone S$. 
Projection is a unary operator
$\pi_{X}(\joinone S)$ that restricts tuples of $\joinone S$ to attributes in $X$. 
Selection is a unary operator $\sigma_\varphi(\joinone S)$
where $\varphi$ is a propositional formula. The operator selects all tuples in the relation 
$\joinone S$ for which $\varphi$ holds.

Consider the movie database in Figure~\ref{fig:toyex1}.
A possible query is: select drama movies and project movie and age of its director. 
We can write this query as follows, 
\begin{equation*}
q_1 = \pi_{\text{Movie,Age}}(\sigma_{\text{Genre = Drama}}(\mathrm{GM} \jointwo \mathrm{MD} \jointwo \mathrm{DA}))
\end{equation*}
The result of query $q_1$ is a set of pairs: $\{(m_3,30),(m_4,30),$ 
$(m_5,30),(m_6,60),(m_7,60)\}$. Another very similar query is: select drama movies and project
age only. That is, 
\begin{equation*}
q_2 = \pi_{\text{Age}}(\sigma_{\text{Genre = Drama}}(\mathrm{GM} \jointwo \mathrm{MD} \jointwo \mathrm{DA}))
\end{equation*}
Query $q_2$ returns:$\{30,60\}$. Although queries $q_1$ and $q_2$ are
very similar, the projection made by $q_2$ on only Age, has eliminated repeated values.
The results of query $q_1$ tell us how many paths there are between directors of Drama and Age,
while in query $q_2$ we only know if a path exists or not.

Our goal is to assess whether the results returned 
by a query provide significant information about our hypothesis on the data. 
For simplicity, a {\it statistic} $f$ is required to map the results
of a query to a single real value. We assume this function $f$ 
is provided by the user together with the query; 
they define
the hypothesis on the data the user wants to test. 
Examples of this statistic are the average of the returned results, or the
number of tuples in the answer, but indeed $f$ can be any general function
returning a real value.

For example, the average value of Age in query 
$q_1$ is 42.5 (i.e., the average age of directors of Drama weighted
by the number of directed movies). 
Then, we may want to know whether that average age 
is interesting or not.
Another two-tailed hypothesis is whether that average 
is significantly different from the average age of directors of romance movies. 

Formally, our problem reads as follows.

\begin{problem}
Given a set of 
binary relations $\{A_1,$ $\ldots,A_n\}$ 
of structured data
and a query
$q$ on some occurring $S \subseteq \{A_1,$ $\ldots,A_n\}$, is the value of $f(q(\joinone S))$ 
for a statistic $f$, significant (in some sense to be made more specific later)? 
\end{problem}

\section{Overview of the method}

In this section we present an overview of the approach and describe the 
intuition behind it. 
We show how our method can be used to 
test the significance of queries and to uncover the structurally
important relations in the data. 

\subsection{Significance testing via randomizations}

We approach the problem of testing the statistical significance of the
query via randomizations. 

Randomizations have been widely used as a method to generate samples from 
null distributions. For example, in medical studies it is customary to measure the effect 
of a certain drug via permutation tests between the control 
and case group~\cite{Good:2000}. 

For short, let $R = \joinone S$ for some $S \subseteq \{A_1,\ldots,A_n\}$.
To assess the significance of $f(q(R))$,
we generate randomized versions of $R$ and run the same query over the samples. 
Let 
$\hat{\mathcal{R}} = \{\hat{R}_1,\ldots,\hat{R}_k\}$ be a set of randomizations of $R$. 
We will specify in Section~\ref{subsec:randtypes} how to generate such randomized versions of $R$. 
Then the one-tailed {\it empirical $p$-value} of $f(q(R))$ with the 
hypothesis of $f(q(R))$ being small is,
\begin{equation}
\frac{|\{\hat{R} \in \hat{\mathcal{R}}: f(q(\hat{R})) \leq f(q(R))\}|+1}{k+1}.
\label{eq:emppvalue}
\end{equation}
This definition represents the fraction of randomized samples having a smaller value of the statistic $f$. 
If the $p$-value is 
small, e.g., below a threshold value $\alpha=0.05$, we can say that the value of $f(q(R))$ is significant in the original 
data.
The one-tailed $p$-value with the hypothesis of $f$ being large and
the two-tailed $p$-value are defined similarly. 
 
\subsection{Where to randomize?}
\label{subsec:whrtorand}

The challenge is how to generate the set $\hat{\mathcal{R}}$, that is,
the different randomized versions of 
$R=\joinone S$, to compute the empirical $p$-value. 
Consider the toy example in Figure~\ref{fig:toyex1}. Suppose we want to evaluate
whether the average age of the directors of drama movies, as in query $q_2$ of Section~\ref{sect:probdef}, 
is young. 
A first naive approach is to consider randomizing directly the binary 
matrix obtained from the boolean product of all relations from Genre to Age. 
The boolean product tells us whether there is a path from the set of nodes
of Genre to the set of nodes of Age, 
as required by query $q_2$.

\begin{figure}
\centering
\subfigure[$\mathrm{GM}\cdot\mathrm{MD}\cdot\mathrm{DA}$]{
\begin{tabular}{l|cc}
\hline
& $30$ & $60$ \\
\hline
{\small Romance} & 1 & 0 \\ 
{\small Drama} & 1 & 1 \\ 
{\small History} & 0 & 1 \\ 
\hline
\end{tabular}}\quad \quad
\subfigure[$\mathrm{GM}\ast\mathrm{MD}\ast\mathrm{DA}$]{
\begin{tabular}{l|cc}
\hline
& $30$ & $60$ \\
\hline
{\small Romance} & 2 & 0 \\ 
{\small Drama} & 3 & 2 \\ 
{\small History} & 0 & 2 \\ 
\hline
\end{tabular}}
\caption{(a) Binary relation $\mathrm{Genre} \times \mathrm{Age}$ obtained via boolean product between
$\mathrm{GM}\cdot\mathrm{MD}\cdot\mathrm{DA}$ of Figure~\ref{fig:toyex1}; (b) Contingency table of paths
between $\mathrm{Genre}$ and $\mathrm{Age}$ obtained via matrix product of $\mathrm{GM}\ast\mathrm{MD}\ast\mathrm{DA}$.}
\label{fig:booleanprod}
\end{figure}

A traditional permutation test\footnote{A traditional permutation test
would swap any values in the matrix, while keeping the row and column sums fixed. 
In binary data this is called swap randomization.} 
on this new matrix shown in Figure~\ref{fig:booleanprod}(a)
can produce only two possible 
random samples: either the original matrix, or a matrix where the age values 
between Romance and History are swapped. For the particular case of romance movies with
the hypothesis of having small age, we would obtain a $p$-value close to 0.5 (i.e.\ 50\% of
the randomized samples would have the same value as the original).
Thus the result is not significant.
Indeed, under such randomization none of the three genres would test significantly 
small, nor large, nor different. 

Alternatively, we could 
apply 
a 
permutation test
on the contingency table of paths~\cite{art/ChenDHL05}, shown in 
Figure~\ref{fig:booleanprod}(b). This table 
gives 
the number of paths between
the Genre and Age, as required by $q_1$. 
The hypothesis related to our queries 
under those permutation tests would never be significant.

The problem of these naive approaches is that they ignore the structure of the 
relations occurring in the query. In our toy example there are three
binary relations participating in the query: GM, MD and DA.  
Indeed these relationships convey some structure on the data:
the relation MD shows that all history movies are also drama movies; the relation MD shows that all movies from Drama and Romance have
been directed by the same person. How do these structures affect the significance of
the results in a query?

In queries involving multiple binary relations, there is
no unique way to randomize. 
To assess 
the structural effect that each 
relation from $S$
has over the query $q(\joinone S)$, we should randomize only the corresponding 
relation.
That is, the different randomizations of $\joinone S$ are obtained by randomizing a single relation $A \in S$ while keeping the rest fixed.

More formally, the random samples of $\joinone S$, when only $A \in S$ is randomized,
are defined as follows:
\begin{equation*}
\hat{\mathcal{R}}_A = \{\joinone\text{T}\cup \hat{A}\ |\ \hat{A} \in \hat{\mathcal{A}} \text{ and } T = S\backslash A\},
\end{equation*}
where $\hat{\mathcal{A}} = \{\hat{A}_1,\ldots,\hat{A}_k\}$
is the set of randomized versions of the original $A \in S$. In Section~\ref{subsec:randtypes} 
we describe the different randomization techniques to obtain such samples. 
Finally, these randomized samples $\hat{\mathcal{R}}_A$ will be used to compute the 
corresponding $p$-value, as described in Equation~\ref{eq:emppvalue}. 

Observe that for a given query involving relations in $S$, 
we can obtain one $p$-value for each $A \in S$ we randomize (while keeping $S\backslash A$
fixed). Each $p$-value is interesting as it measures the structural effect that the participant
relation $A$ has on the significance of the result of the query. 

The sketch of the method is described in Algorithm~\ref{alg:method}. The basis of our proposal 
can be found in traditional statistics under the name of restricted randomizations (see e.g.~\cite{Good:2000},
typically to test whether a treatment variable has effect on a response variable). 

\begin{algorithm}[t]
\caption{Query significance in multi-relational data}
\begin{algorithmic}[1]
\Input  A set of binary relations $S \subseteq \{A_1,\ldots,A_n\}$,
a que\-ry $q(\joinone S)$ and a hypothesis over the statistic $f(q(\joinone S))$
\Output A set of $p$-values
\For{{\bf each} binary relation $A \in S$}
\State Obtain $k$ random samples of $A$, $\hat{\mathcal{A}} = \{\hat{A}_1,\ldots,\hat{A}_k\}$\label{alg:qs:obt}
\State Let $\hat{\mathcal{R}}_A = \{\joinone\text{T}\cup \hat{A}\ |\ \hat{A} \in \hat{\mathcal{A}} \text{ and } T = S\backslash A\}$\label{alg:qs:comb}
\State Compute the $p$-value using the random samples $\hat{\mathcal{R}}_A$
\EndFor
\end{algorithmic}
\label{alg:method}
\end{algorithm}

\subsection{Example}
\label{sub:example}
We study now the toy example in Figure~\ref{fig:toyex1}.
Consider a query defined such as $q_1$ from Section~\ref{sect:probdef},
yet on the three different Genres. 

The first hypothesis that romance movies are directed by young directors 
obtains a $p$-value of 0.131 when randomizing on GM, 
a $p$-value of 0.494 on MD, and a $p$-value of 0.495 on MA.
The hypothesis is not significant under any randomization, 
but we observe that randomizing on GM obtains the smallest $p$-value 
for this query. 

The hypothesis that history movies are directed by old directors 
obtains $p$-values 0.269, 0.045, 0.495 when randomizing on GM, MD, DA, respectively. 
Thus the hypothesis is significant considering the structure 
in relation MD: all
non-history movies are directed by the same person.

Finally, the hypothesis that drama movies are directed by young 
directors 
is not significant in any of the randomizations, always with a $p$-value close to $1$
when randomizing on GM or MD, and $p$-value of 0.495 when randomizing on DA.

In summary: the age value of 30 associated to romance movies
is close to being significant when randomizing on GM because 
Romance is a non-intersecting genre with Drama and History; 
the age value of 60 associated to history movies
is significant when randomizing on MD because, when focusing on the directors,
the history movies are non-intersecting
with the romance and drama movies---all romance and drama movies are directed
by the same person; also, the relation DA always swaps
with equal probability, because of its one-to-one structure. In the next section we will understand 
better the reason of these explanations.

\section{Randomizations in multi-re\-la\-ti\-onal model}

This section describes how to obtain random samples for a single
relation $A$ (line~\ref{alg:qs:obt} in Algorithm~\ref{alg:method}), and presents the combinatorial
properties of combining such samples with the other relations in
the query (line~\ref{alg:qs:comb} in Algorithm~\ref{alg:method}). 

\subsection{Types of randomization}\label{subsec:randtypes}

Given a binary relation $A$ we use three different types of randomization to
obtain random samples from $A$. The running times and space consumptions of the methods are linear in the size of the relation $A$.
\begin{itemize}
\item[(1)] {\it Swap randomization} 
of $A$, as used in~\cite{cobb03,swaprand-gionisetal},
produces random samples of $A$ that preserve 
the row and column sums. 
The algorithm starts from $A$ 
and performs 
local swaps interchanging a pair of 1's with a pair of 0's 
preserving the row and column sums. 
Technically, a local swap consists of selecting entries $(i,j),(k,l) \in A$ such that
$(i,l),(k,j) \notin A$, and swapping the elements so that $(i,j),(k,l) \notin A$ and
$(i,l),(k,j) \in A$. On the bipartite graph representation of the relation $A$,
a local swap represents a flip between two independent edges. 
\begin{center}
\includegraphics[width=2.3cm]{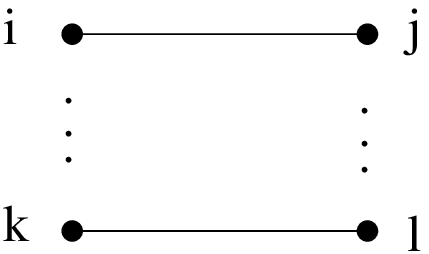}
\quad
\raisebox{5.5mm}{$\iff$}
\quad
\includegraphics[width=2.3cm]{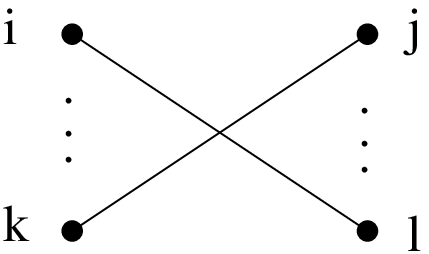}
\end{center}
A sequence of swaps is performed until the data mixes
sufficiently enough in a Markov chain approach~\cite{besag,besagclifford89}, and therefore, a random sample 
of $A$ is obtained. We use ten times the number of ones in the matrix as the number of swaps, which suffices for the convergence of the chain~\cite{swaprand-gionisetal}. We denote the set 
of all random samples reached via swap randomization of $A$ 
as $\sw(A)$.
\item[(2)] {\it Row permutation} 
of $A$ 
permutes 
the order of
the rows of $A$. We denote the set of all random
samples 
reached 
via row permutation of $A$ as $\rp(A)$.
\item[(3)] {\it Column permutation} 
of $A$ 
permutes 
the order of the columns of $A$. We denote the set of all random
samples 
reached 
via column permutation of $A$ as $\cp(A)$.
\end{itemize}
Note, particularly, that $\sw(A)$, $\rp(A)$ and $\cp(A)$ refer to sets of matrices.
The relationship between swap randomizations and permutations can be stated as follows.

\begin{proposition} Let $A$ be a binary matrix. Then:
\begin{itemize}
\item $\rp(A) =\sw(I) \cdot A$, where $I$ is an identity matrix; 
\item $\cp(A) = A \cdot \sw(I)$, where $I$ is an identity matrix;
\item if $A$ has one 1 in each row, then $\sw(A) = \rp(A)$;
if $A$ has one 1 in each column, then $\sw(A) = \cp(A)$.
\end{itemize}
\label{prop0}
\end{proposition}

Note that $\sw(I)$, for identity matrix $I$, can produce any swap permutation matrix 
with uniform distribution. Thus, we have that the boolean product $\sw(I) \cdot A$ produces all permutations for the rows of $A$
and similarly, 
$A \cdot \sw(I)$ produces all permutations of the columns of $A$. 
Intuitively, these row (or column) permutations can be seen as a random re-assignment
of the row (or column) names in $A$. 

While the swap randomization has been used in~\cite{swaprand-gionisetal}
to assess the data mining results on a single binary relation, the 
new randomizations, corresponding to row 
and column permutations,
do not make sense in such a context. The row
or column permutation of a matrix does not change any of the 
frequent pattern solutions in the new randomized matrix. 
These permutations only make sense in a multi-relational
data model, where the permuted matrices are combined with other relations. 
Both row and column permutation
of a single relation change the global paths from the source nodes 
to  destination nodes in the query graph, and thus, the evaluation of the query 
can change on the randomized data.

\subsection{Properties}

Next we study the properties of combining the obtained random samples
with the other relations in the query. 
For simplicity, we study the case of queries with only two occurring relations $q(A\jointwo B)$ 
and use boolean product as a simplification of the natural join. 
For notational convenience, we overload the boolean product for the sets of 
binary matrices, e.g., $\sw(A) \cdot \sw(B)$
represents the boolean product of each pair of elements $A \in \sw(A)$ and $B\in \sw(B)$.

The following 
inclusions with swap randomization follow immediately after
the definitions. All other inclusions 
do not hold. The inclusions
can also be proper in all cases. 

\begin{proposition} Let $A,B$ be binary matrices. Then:
\begin{itemize}
\item $A \cdot B \subseteq \sw(A) \cdot B \subseteq \sw(A) \cdot \sw(B)$; 
\item $A \cdot B \subseteq A \cdot \sw(B) \subseteq \sw(A) \cdot \sw(B)$;  
\item $A \cdot B \subseteq \sw(A \cdot B)$.
\end{itemize}
\label{prop1}
\end{proposition}

Proposition~\ref{prop1} tells us that the set of samples that can
be obtained by randomizing two relations is larger than by randomizing
only one relation. As discussed in Section~\ref{subsec:whrtorand}, we prefer to randomize
a single table at a time in order to control much better the structural
effect the randomized relation has on the query. Additionally, we know that 
the set of randomized samples $\sw(A) \cdot B$ is different from the set $A \cdot \sw(B)$,
thus it makes sense to do them both separately.  

Next we present several properties relating swap randomization to
row and column permutations. 

\begin{proposition} 
Let $A,B$ be binary relations. 
If $B$ is a one-to-one relation, then $A \cdot \sw(B) = \cp(A)$. 
If $A$ is a one-to-one relation, then $\sw(A) \cdot B = \rp(B)$. 
\label{prop2}
\end{proposition}

Proposition~\ref{prop2} follows immediately
from Proposition~\ref{prop0}. In real world datasets, it is quite common to have one-to-one
relations. For example, 
the ages of the directors in the example in Figure~\ref{fig:toyex1} are one-to-one. Thus swap randomization of the relation DA produces the same set of samples as the column permutation of MD.

\begin{proposition} Let $A,B$ binary relations. Then:
\begin{itemize}
\item $\cp(A \cdot B) = A \cdot \cp(B)$
\item $\rp(A \cdot B) = \rp(A) \cdot B$
\item $\cp(A) \cdot B = A \cdot \rp(B) = \cp(A) \cdot \rp(B)$
\end{itemize}
\label{prop3}
\end{proposition}
This means that column and row permutations do not make sense in more than one relation, e.g., $A \cdot \cp(B \cdot C \cdot D) \cdot E = A \cdot B \cdot C \cdot \cp(D) \cdot E$. 
The last property of Proposition~\ref{prop3} states 
that only one permutation, either column permutation
on $A$ or row permutation on $B$, is indeed necessary. 

Finally, we give an implication of Proposition~\ref{prop0} that reduces the number of 
different randomizations considerably.
\begin{theorem} Let $A,B$ be binary relations. Then:
$A \cdot \sw(I) \cdot B = \cp(A)\cdot B = A \cdot \rp(B)$, where $I$ is
an identity matrix.
\label{prop4}
\end{theorem}
Hence, we prefer to use the notation with the identity matrix $I$ 
to refer to the row and column permutations. 
The operation $A \cdot \sw(I) \cdot B$ randomizes the boolean product,
whereas the operations $\sw(A) \cdot B$ and $A \cdot \sw(B)$ randomize the original
data. From this perspective, $A \cdot \sw(I) \cdot B$ tells about the significance of the combination operation, while $\sw(A) \cdot B$ tells whether the
structure in $A$ is significant. To sum up, we have the following result:
\begin{corollary}
For a query $q(A \jointwo B)$, 
there exist three different randomizations: 
(i) $\sw(A)$ while keeping $B$ fixed; 
(ii) $\sw(B)$ while keeping $A$ fixed; 
(iii) $\sw(I)$ where $I$ is an identity 
relation between the columns of $A$ and the rows of $B$.
\end{corollary}
Notice that if $A$ or $B$ are one-to-one relations, then randomization (iii) 
will be the same as (i) or (ii) respectively. 
Each randomization provides a set of
samples from where we can compute a $p$-value for our query (hypothesis on the data). 
Every $p$-value is interesting as it shows how the structure of the randomized relation affects 
the significance.

\subsection{Example revisited}
The $p$-values reported in Section~\ref{sub:example} for the toy example in Figure~\ref{fig:toyex1},
correspond to swap randomization of the binary tables GM, or MD, or DA, respectively. 
Indeed because MD has one single 1 in each row, we have that 
$\text{GM} \cdot \sw(I) \cdot \text{MD} \cdot \text{DA}$ is equal to 
$\text{GM} \cdot \sw(\text{MD}) \cdot \text{DA}$. Similarly, because
DA is a one-to-one relation, we have
$\text{GM} \cdot \text{MD} \cdot \sw(I) \cdot \text{DA}$ equals
$\text{GM} \cdot \text{MD} \cdot \sw(\text{DA})$.
Thus, for this example, only swap randomization in the three tables is necessary. 

Interestingly, we can understand better now the $p$-values reported in Section~\ref{sub:example}. 
On the relation GM, drama movies and history movies have no 
independent edges to swap between them.
Therefore, the pattern of History implying 
Drama tends to remain in random samples. As a result, the $p$-value
of the hypothesis related to history or drama movies is not significant. On the other hand,
the $p$-value related to romance movies becomes close to being significant because, for this genre,
the null distribution diverges more from the original. The fact that there are only
two romance movies raises this $p$-value slightly above the 0.05 threshold.

Similar explanation goes when randomizing MD. When looking at MD,
local swaps can interchange at most two edges between movies of the young director C.~Waitt 
and movies of the not-so-young director T.~George. 
Actually, in all random samples coming from MD we observe that 
C.~Waitt has always at least three movies from either drama or romance. 
As a result, neither drama nor romance can be significant---in the null distribution they
are always closely linked to a young director as in the original data. Yet,
history movies directed
by T.~George have more local swaps that would create a diverging null distribution---most
of the samples in the null distribution have the history movies connected to the age of 30.
The hypothesis of history movies being directed by a not-so-young person is then significant.

\section{Studying path distributions}

For a query $q(A \jointwo \ldots \jointwo B)$ where 
$A \subseteq I \times L$ and $B \subseteq J \times K$, 
let $P = A \ast \ldots \ast B$ 
be the matrix product of all relations participating in $q$. This
corresponds to the contingency table of paths from origin $I$ to destination nodes $K$.
An example is shown in Figure~\ref{fig:booleanprod}(b) for the toy data of Figure~\ref{fig:toyex1}.
For all types of queries, the significance of the result is closely related 
to the path distributions between nodes $I$ and $K$. 
For example, suppose we want to test whether the average age of history-movie
directors is large. In the original data of Figure~\ref{fig:toyex1:graph} 
there are two paths from History to the age of 60 and no path to the age of 30. 
It is sensible to assume that if we had random samples where paths are mainly 
swapped the other way round, the hypothesis would be significant. 

Naturally, a simple way to visualize whether there exists an interesting
finding in the data is to compare the path distribution of $P$ with 
the expected path distribution on the given random samples. The larger the 
change, the more significant the result would tend to be. 

The following three matrices
show the expectation of the paths when swap randomizing relation GM, MD or DA, respectively,
for the example in Figure~\ref{fig:toyex1}. 
\\[.5em]
{\scriptsize
$E[\sw(\text{GM}) \ast \text{MD} \ast \text{DA}]$ 
\hspace{\stretch{1}}
$E[\text{GM} \ast \sw(\text{MD}) \ast \text{DA}]$
\hspace{\stretch{1}}
$E[\text{GM} \ast \text{MD} \ast \sw(\text{DA})]$
} 
\\[.5em]
{\small
$\begin{pmatrix}
\mathbf{0.849}   & \mathbf{1.151}\\ 3.269  & 1.731 \\ 0.882 & 1.118
\end{pmatrix}$
\hspace{\stretch{1}}
$\begin{pmatrix}
1.413 & 0.587 \\ 3.587 & 1.413 \\ \mathbf{1.455} & \mathbf{0.545}
\end{pmatrix}$
\hspace{\stretch{1}}
$\begin{pmatrix}
0.984 &    1.016 \\ 2.492 & 2.508 \\ 1.016 & 0.984
\end{pmatrix}$
\hspace{-2mm}\\
}

\looseness-1
The genre that swaps most of its paths under randomizations with GM is Romance. 
History swaps the paths
from the age of 60 to the age of 30 when randomizing on MD. 
Randomization on DA distributes paths fifty-fifty for each
genre. The $p$-values obtained there were always close to 0.5.

\section{Empirical results}

In this section we present empirical results on synthetic and real datasets. 
Our real dataset is \textbf{MovieLens}, which is very similar to IMDb.
In all cases, we calculate the empirical $p$-values over 999 randomized samples and use the threshold of $\alpha =0.05$ to determine the query significance. 

The randomization methods are fast in practice. In our experiments, producing one randomized sample took approximately the same time as evaluating the query. With the tested datasets, the times for producing one sample were at most few seconds with Java implementations integrated with MATLAB on a 2.2GHz Opteron. 
The time and space consumption of the methods scale linearly in the size of the relation. In large-scale applications, fewer number of randomized samples can be used to calculate the empirical $p$-values. For example, 30 samples is usually sufficient in a preliminary significance analysis. This corresponds to approximately 30 times increase in the evaluation time.

\subsection{Synthetic dataset}

To motivate our approach and understand better why randomizations are consistent with
the inferences about our hypothesis, we generate a synthetic dataset to simulate relations of
users, movies and genres. We will be interested in testing the following hypothesis.
\begin{query}\label{query:synth:mldtomtw}
Men watch different types of movies than women.
\end{query}
The relations occurring in the query are: 
Gender$\times$User (SU), User$\times$Movie (UM) and Movie$\times$Genre (MG).  

For studying the behavior of randomizations, 
we generate the tables SU, UM and MG to make our hypothesis clearly be significant. 
We let SU contain 30 men and 20 women, thus SU is a $2\times50$ binary table where the first 30 values 
in the first row and the last 20 values in the second row are 1s. We generate UM to 
be a $50\times100$ binary 
table where men watch any of the first 60 movies with probability of 0.40 and any of the last
40 movies with probability of 0.05. To create a strong pattern, we let the probabilities of a female
watching movies be the other way round. 
Finally, we generate MG as a $100\times6$ binary table where the first three genres 
will be considered to be manly and the last three genres will be considered to be womanly. 
For each movie in the relation, we select two genres as follows: 
for the first 60 movies we select a genre from the manly genres with a probability of 0.9 
and from the womanly genres with a probability of 0.1. For the last 40 movies the probabilities are 
the other way round. So, each movie has at most two genres, because if we happen to 
select the same genre for a movie twice, then we say that the movie has only one genre.

Next we create the anti-tables from those above, called rSU, rUM and rMG. 
These anti-tables will not contain any structure at all, they are random.
We let rSU be a $2\times50$ binary table with 30 men and 20 women where the order of the users is random. 
We generate rUM to be a $50\times100$ table with each element being 1 with a probability of $(0.40+0.05)/2$. 
And we let rMG be formed similarly to MG but with the two genres for each movie assigned 
uniformly with replacement. 

\begin{figure}[t]
\subfigure[SU*UM*MG]{\includegraphics[width=0.22\textwidth]{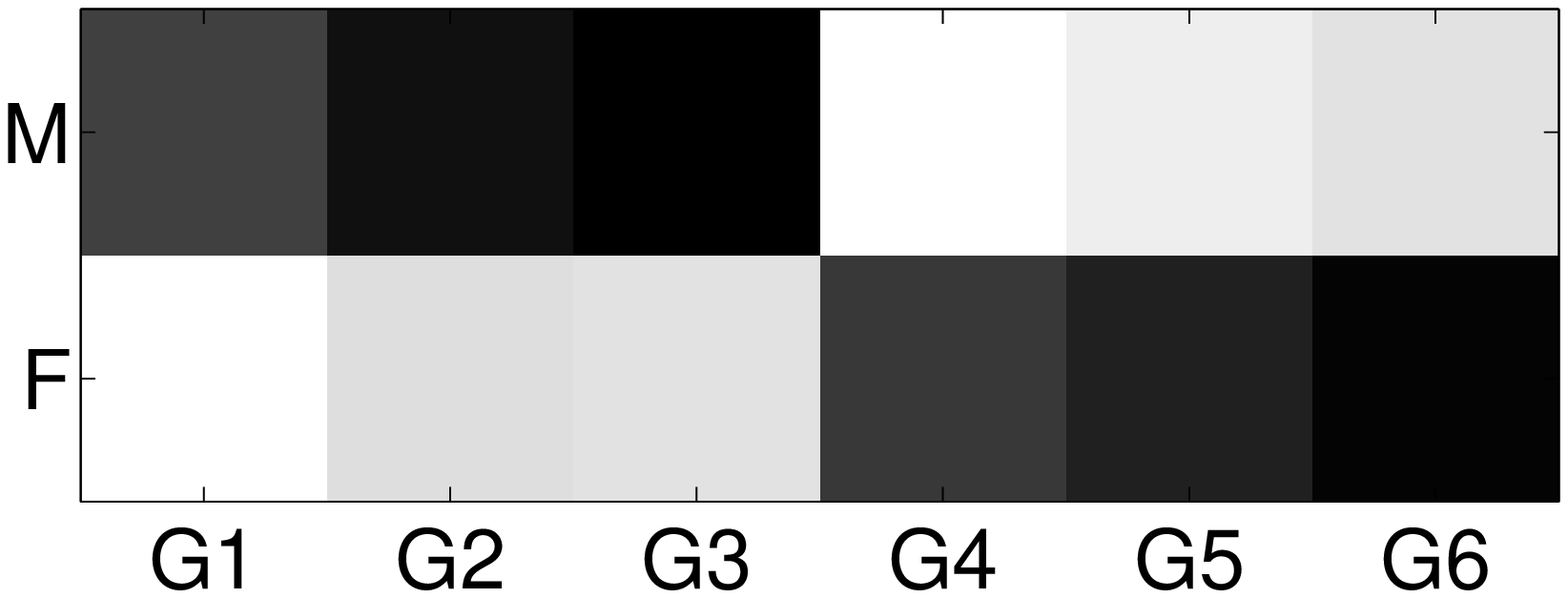}}
\hspace{\stretch{1}}
\subfigure[rSU*UM*MG]{\includegraphics[width=0.22\textwidth]{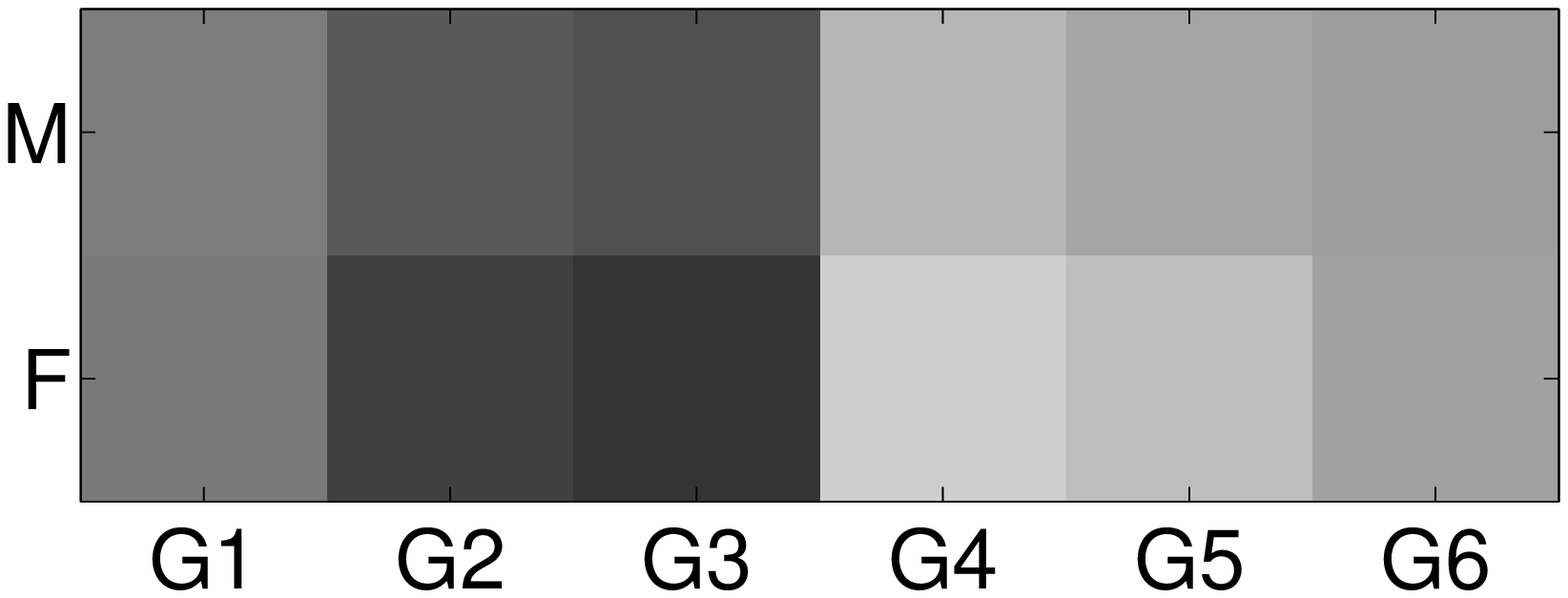}}
\subfigure[SU*rUM*MG]{\includegraphics[width=0.22\textwidth]{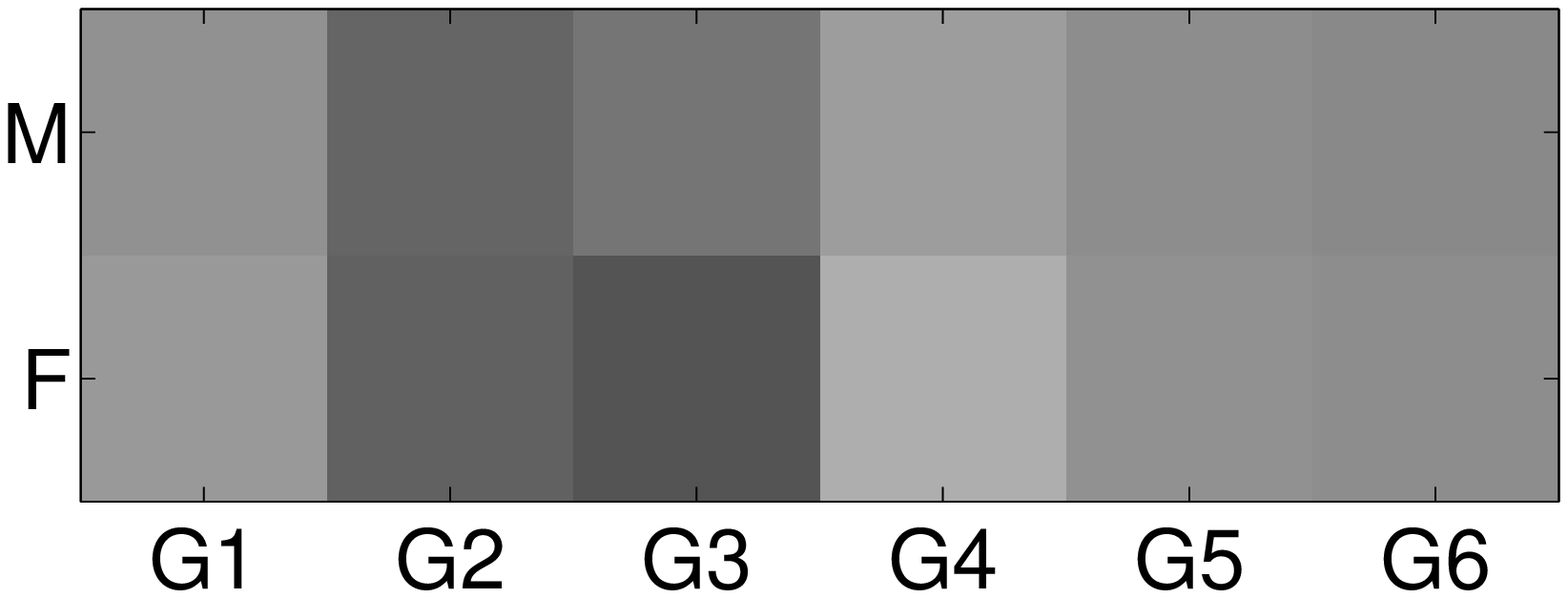}}
\hspace{\stretch{1}}
\subfigure[SU*UM*rMG]{\includegraphics[width=0.22\textwidth]{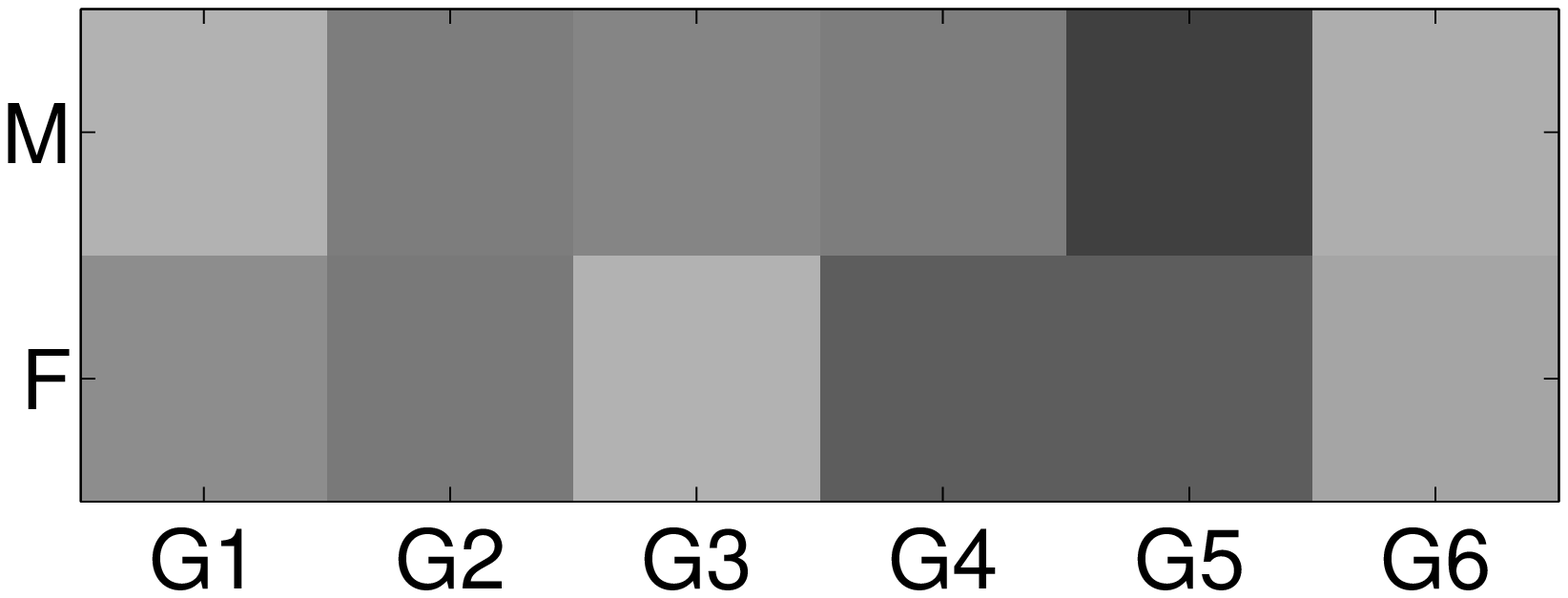}}\hspace{-0.9mm}
\caption{Proportion of paths going from a gender (M=male, F=female) 
to a genre (G1--G6) in the different combined tables. Lighter color represents 
less paths, while darker more paths; to be more exact: white corresponds to the lowest value of 
4.5\% and black to the highest value of 30\%.}
\label{fig:synt:paths:distr}
\end{figure}

The goal of this experiment is to study how the $p$-values of Hyp~\ref{query:synth:mldtomtw} 
change when combining the original significant tables SU, UM and MG to 
one of these non-significant tables. Figure~\ref{fig:synt:paths:distr} shows
the contingency table of paths from those combinations. 
We notice that using the original tables SU, UM and MG (Figure~\ref{fig:synt:paths:distr}(a))
produces clearly a significant difference between the types of movies that males and females watch. 
By replacing one of the original tables with a random version, the pattern seems to disappear. 
Still, we cannot clearly see from the path distributions which of the underlying tables mainly 
breaks the original structure. We would like to check with our tests whether randomizing in 
the proper tables will tell us where the pattern is broken.

For the test, we use the following statistic.
\begin{stat}\label{stat:synth:mldtomtw}
$L_1$ distance between the distribution of genres of the movies that men and women have watched.
\end{stat}
This statistic is the sum of the absolute differences between the proportion of paths of men and women, 
as shown in Figure~\ref{fig:synt:paths:distr} for each of the combinations. 
The original value of the statistic with the tables SU, UM and MG is 1.23, implying a clear 
difference between males and females. When one of the tables SU, UM and MG is replaced with a corresponding anti-table, the value of the $L_1$ statistic is around 0.1. 

\begin{table}
\small
\centering
\begin{tabular}{c*{2}{@{\hspace{3mm}}c}*{4}{r}} 
\toprule
\multicolumn{3}{c}{Input relations} & \multicolumn{4}{c}{$p$-values} \\ 
\cmidrule(lr){1-3} \cmidrule(lr){4-7}
A & B& C & \sw($I_{AB}$) & \sw(B) & \sw($I_{BC}$) & \sw(C) \\
\midrule
SU & UM & MG & 0.001 & 0.001 & 0.001 & 0.001 \\
rSU & UM & MG & {\bf 0.517} & 0.030 & 0.013 & 0.003 \\
SU & rUM & MG & {\bf 0.282} & {\bf 0.279} & {\bf 0.155} & 0.124 \\
SU & UM & rMG & 0.001 & 0.001 & {\bf 0.704} & {\bf 0.727} \\
\bottomrule
\end{tabular}
\caption{Significance tests for the Hyp~\ref{query:synth:mldtomtw} with the combined input relations $\mathrm{A}\jointwo\mathrm{B}\jointwo\mathrm{C}$. 
The first three columns contain the relations considered as input, labeled $\mathrm{A}$, $\mathrm{B}$ and $\mathrm{C}$. 
Columns 4th to 7th are empirical $p$-values for 
the hypothesis when only one relation is randomized: $\sw(I_{AB})$ randomizes the identity matrix between relations $\mathrm{A}$ and $\mathrm{B}$, which is equivalent to randomizing the relation $\mathrm{A}$, $\sw(\mathrm{A})$; 
$\sw(\mathrm{B})$ randomizes only on relation $\mathrm{B}$; $\sw(I_{BC})$ randomizes the identity matrix between
relations $\mathrm{B}$ and $\mathrm{C}$; $\sw(\mathrm{C})$ randomizes only relation 
$\mathrm{C}$. 
Bold $p$-values correspond to randomizations which touch the anti-tables.
}
\label{table:sign:synth}
\end{table}

In Table~\ref{table:sign:synth} we show the results of the several significance tests
for the hypothesis Hyp~\ref{query:synth:mldtomtw} on the several combined tables. 
There is a clear connection between the structure of the relations A, B and C occurring in the query 
and the $p$-values obtained by randomizing in different relations. 
As expected, the empirical $p$-value of Hyp~\ref{query:synth:mldtomtw} with tables 
SU, UM and MG is significant with randomizations in all tables. 
On the other hand, when one of 
the clearly-structured tables SU, UM or MG 
is replaced by the anti-tables rSU, rUM or rMG respectively, 
we obtain large empirical $p$-values for those randomizations 
that touch the anti-tables (see the bold values of Table~\ref{table:sign:synth}).
This illustrates how randomizations can tell about the structural
effects in the significance of a query.

\subsection{MovieLens dataset}

The \textbf{MovieLens} data is collected through the MovieLens web site
(\url{movielens.umn.edu}). 
The downloadable data is already cleaned up, i.e., 
users who had less than 20 ratings or did not have complete demographic 
information were removed from the data set. In all, the data consists of
100,000 ratings (valued from 1 to 5) from 943 users on 1,682 movies. Each 
user has rated at least 20 movies and the demographic information for the
users correspond to attributes of age, gender, occupation and zip code. 
For each movie we have title, release year and a list of genres. 
Furthermore, we interpret that if a user has rated a movie, it means that
he or she has watched it. This corresponds to the binary table named UM. We do not use the information of ratings in any other way. 
In Table~\ref{table:movielens:summary} we summarize the binary relations in 
the \textbf{MovieLens} dataset. The table UA is just an identity matrix which maps the users to their ages, thus two different columns of the table UA may correspond to the same age. Handling numerical values in this way guarantees that two users having the same age are not combined into a single user after a join and a projection.

\begin{table}
\small
\centering
\begin{tabular}{l@{\hspace{4mm}}lrrr}
\toprule
Relation & Description & Rows & Cols & \# of 1's/row \\
\midrule
UM & User$\times$Movie & 943 & 1680 & 106 \\
MG & Movie$\times$Genre & 1680 & 18 & 1.7 \\
UO & User$\times$Occupation & 943 & 21 & 1 \\
US & User$\times$Gender & 943 & 2 & 1 \\
UA & User$\times$Age & 943 & 943 & 1 \\
\bottomrule
\end{tabular}
\caption{Summary of tables in MovieLens dataset. 
The table $\mathrm{UA}$ is an identity map between users and their ages.
We denote a transpose by reversing the relation name.}
\label{table:movielens:summary}
\end{table}

Next, we go through a few queries on the dataset and analyze their significances.

\begin{query}\label{query:mldtomtw}
Men watch different types of movies than women.
\end{query}

\begin{stat}\label{stat:mldtomtw}
$L_1$ distance between the distribution of genres of the movies that men and women have watched.
\end{stat}

In Table~\ref{table:sign:query:mldtomtw} we give the empirical $p$-values for Hyp~\ref{query:mldtomtw}. Each
row shows the relation being randomized for obtaining the corresponding $p$-value. 
The query associated to the hypothesis traverses the relations Gender $\times$ User $\times$ Movie $\times$ Genre, 
corresponding to relations SU, UM and MG. There are five different types of randomizations of the query 
which each produce a unique $p$-value. The results in Table~\ref{table:sign:query:mldtomtw} show that 
Hyp~\ref{query:mldtomtw} is significant wrt all different randomizations. 

\begin{table}
\small
\centering
\begin{tabular}{lr@{\hspace{2mm}}rr}
\toprule
 & Mean & (Std) & $p$-value \\
\midrule
$\mathrm{SU} \jointwo \mathrm{UM} \jointwo \mathrm{MG}$ & 0.16 & \\
$\sw(\mathrm{SU}) \jointwo \mathrm{UM} \jointwo \mathrm{MG}$ & 0.03 & (0.01) & 0.001 \\
$\mathrm{SU} \jointwo \sw(I) \jointwo \mathrm{UM} \jointwo \mathrm{MG}$& 0.03 & (0.01) & 0.001 \\
$\mathrm{SU} \jointwo \sw(\mathrm{UM}) \jointwo \mathrm{MG}$& 0.01 & (0.00) & 0.001 \\
$\mathrm{SU} \jointwo \mathrm{UM} \jointwo \sw(I)\jointwo \mathrm{MG}$& 0.03 & (0.01) & 0.001 \\
$\mathrm{SU} \jointwo \mathrm{UM} \jointwo \sw(\mathrm{MG})$& 0.02 & (0.00) & 0.001 \\
\bottomrule
\end{tabular}
\caption{Significance evaluation of Hyp~\ref{query:mldtomtw}. 
Mean and std are the average and standard deviation of Statistic~\ref{stat:mldtomtw} in the original
input data (first row) and several randomizations.}
\label{table:sign:query:mldtomtw}
\end{table}

Indeed, the results on Hyp~\ref{query:mldtomtw} seem to indicate that men watch movies with different 
genres than women. All randomizations are consistent. 
We will next analyze which genres separate men and women. We repeat the following hypothesis (with associated
query) for each genre $G$.
 
\begin{query} \label{query:mlggmtw}
Men watch genre $G$ more (or less) than women.
\end{query}
\begin{stat} \label{stat:mlggmtw}
The difference between the \%-proportions of the movies from genre $G$ among all 
the movies men and women have watched. 
\end{stat}
Notice this statistic is similar to Statistic~\ref{stat:mldtomtw} but now we only look at the difference for the specific
genre $G$. The empirical $p$-values of the significance testings of Hyp~\ref{query:mlggmtw} are given in 
Table~\ref{table:sign:query:mlggmtw}. Again we find out that randomizing in different relations
produces fairly similar results in general. We can observe that men watch 
significantly more, for example, action and sci-fi movies than women,
whereas women watch significantly more romance and drama movies than men.
Interestingly, we can say the popularity of mystery and documentary movies do not depend on the gender. 
Actually the genres which have the smallest amount of movies are the least significant ones. 
The genres with fewest number of movies are fantasy (with 22 movies), film-noir (24), western (27), animation (41) and 
documentary (50).

\begin{table}[t]
\small
\centering
\begin{tabular}{@{\hspace{1.2mm}}l@{\hspace{1mm}}r*{5}{@{\hspace{2mm}}r}@{\hspace{1.2mm}}}
\toprule
$G$ & Orig. & \sw(SU) & \sw($I_1$) & \sw(UM) & \sw($I_2$) & \sw(MG) \\
\midrule
Action & 2.5 & 0.001 & 0.001 & 0.001 & 0.001 & 0.001 \\
Sci-fi & 1.5 & 0.001 & 0.001 & 0.001 & 0.001 & 0.001 \\
Thriller & 1.1 & 0.001 & 0.001 & 0.001 & 0.001 & 0.001 \\
Adventure & 0.8 & 0.001 & 0.001 & 0.001 & 0.001 & 0.001 \\
Crime & 0.6 & 0.002 & 0.001 & 0.001 & 0.001 & 0.002 \\
War & 0.5 & 0.002 & 0.001 & 0.001 & 0.004 & 0.002 \\
Horror & 0.4 & 0.019 & 0.018 & 0.001 & 0.011 & 0.020 \\
Western & 0.2 & 0.001 & 0.001 & 0.001 & 0.005 & 0.003 \\
\hdashline
Film-noir & 0.1 & 0.012 & 0.009 & 0.001 & 0.054 & 0.058 \\
Mystery & 0.0 & 0.392 & 0.401 & 0.395 & 0.424 & 0.469 \\
Document. & 0.0 & 0.404 & 0.392 & 0.391 & 0.468 & 0.489 \\
Fantasy & -0.1 & 0.064 & 0.070 & 0.051 & 0.243 & 0.201 \\
\hdashline
Animation & -0.2 & 0.032 & 0.033 & 0.001 & 0.027 & 0.018 \\
Musical & -0.5 & 0.001 & 0.001 & 0.001 & 0.001 & 0.001 \\
Children's & -1.0 & 0.001 & 0.001 & 0.001 & 0.001 & 0.001 \\
Comedy & -1.3 & 0.001 & 0.001 & 0.001 & 0.001 & 0.001 \\
Drama & -2.3 & 0.001 & 0.001 & 0.001 & 0.001 & 0.001 \\
Romance & -2.3 & 0.001 & 0.001 & 0.001 & 0.001 & 0.001 \\
\bottomrule
\end{tabular}
\caption{Empirical $p$-values for Hyp~\ref{query:mlggmtw}.
The values for the associated Statistic~\ref{stat:mlggmtw} in the original relations are given in the second column. 
The different randomizations methods (columns 3rd to 7th) correspond to randomizing in one relation at a time from 
$\mathrm{SU}\jointwo I_1 \jointwo \mathrm{UM} \jointwo I_2 \jointwo \mathrm{MG}$. Genres are sorted by the value of the statistic. 
Significance tests say: genres over the first dashed line are more watched by men ($p$-values always
under 0.05); genres under the second dotted line are more watched by women ($p$-values always under 0.05). 
We cannot say anything about genres in between the two dotted lines.}
\label{table:sign:query:mlggmtw}
\end{table}

Next we study users by their occupation. 
\begin{query} \label{query:uwagowdtomtou}
The users with occupation $O$ watch different types of movies than other users.
\end{query}
\begin{stat} \label{stat:uwagowdtomtou}
$L_1$ distance between the distributions of genres of the movies watched by users with occupation $O$ and users with other occupations.
\end{stat}
The results of the significance testings are given in Table~\ref{table:sign:query:uwagowdtomtou}. 
When evaluating the associated query, we find that randomizing in different relations matters for that query. 
For most of the occupations, Hyp~\ref{query:uwagowdtomtou} is not significant when randomizing
on $\sw(\mathrm{OU})\jointwo \mathrm{UM}\jointwo \mathrm{MG}$ 
nor $\mathrm{OU}\jointwo \sw(I)\jointwo \mathrm{UM}\jointwo \mathrm{MG}$. For the other
randomizations we have that all occupations, except for homemakers, exhibit 
significance of the hypothesis.  
We observe that the largest occupation groups of librarians (51), educators (95) and students (196) have the
most significant empirical $p$-values for the query, with all type of randomizations. We could infer that those type of users watch different genres than other users.

\begin{table}[t]
\small
\centering
\begin{tabular}{@{\hspace{1mm}}l@{\hspace{1mm}}r*{2}{@{\hspace{3mm}}r@{\hspace{1mm}}r@{\hspace{2.5mm}}r}@{\hspace{1mm}}}
\toprule

 & & \multicolumn{3}{c}{$\sw(\mathrm{OU})$} & \multicolumn{3}{c}{$\sw(I_2)$} \\
\cmidrule(r{3mm}){3-5} \cmidrule(r){6-8} 
 & Orig.  & Mean & (Std) & $p$-val. & Mean & (Std) & $p$-val. \\
\midrule
None & 0.23 & 0.13 & (0.05) & \textbf{0.038} & 0.07 & (0.01) & 0.001 \\
Librarian & 0.18 & 0.05 & (0.02) & \textbf{0.001} & 0.04 & (0.01) & 0.001 \\
Retired & 0.18 & 0.10 & (0.04) & \textbf{0.040} & 0.05 & (0.01) & 0.001 \\
Homemaker & 0.17 & 0.14 & (0.05) & 0.269 & 0.15 & (0.03) & \textbf{0.226}\\
Doctor & 0.15 & 0.14 & (0.05) & 0.373 & 0.08 & (0.02) & 0.001 \\
Entert. & 0.14 & 0.09 & (0.03) & 0.073 & 0.04 & (0.01) & 0.001 \\
Educator & 0.13 & 0.04 & (0.01) & \textbf{0.001} & 0.03 & (0.01) & 0.001 \\
Lawyer & 0.13 & 0.11 & (0.04) & 0.237 & 0.05 & (0.01) & 0.001 \\
Salesman & 0.12 & 0.11 & (0.04) & 0.330 & 0.06 & (0.01) & 0.001 \\
Healthcare & 0.12 & 0.09 & (0.03) & 0.211 & 0.04 & (0.01) & 0.001 \\
Student & 0.11 & 0.03 & (0.01) & \textbf{0.001} & 0.03 & (0.01) & 0.001 \\
Scientist & 0.11 & 0.07 & (0.02) & 0.052 & 0.05 & (0.01) & 0.001 \\
Artist & 0.10 & 0.07 & (0.03) & 0.130 & 0.04 & (0.01) & 0.001 \\
Technician & 0.10 & 0.07 & (0.03) & 0.183 & 0.03 & (0.01) & 0.001 \\
Programmer & 0.08 & 0.05 & (0.02) & \textbf{0.025} & 0.03 & (0.01) & 0.001 \\
Engineer & 0.08 & 0.05 & (0.02) & \textbf{0.034} & 0.03 & (0.01) & 0.001 \\
Marketing & 0.08 & 0.07 & (0.03) & 0.340 & 0.05 & (0.01) & 0.006 \\
Writer & 0.08 & 0.06 & (0.02) & 0.122 & 0.03 & (0.01) & 0.001 \\
Executive & 0.07 & 0.07 & (0.02) & 0.337 & 0.04 & (0.01) & 0.001 \\
Administr. & 0.05 & 0.04 & (0.02) & 0.367 & 0.02 & (0.01) & 0.001 \\
Other & 0.04 & 0.04 & (0.01) & 0.483 & 0.02 & (0.00) & 0.002 \\
\bottomrule
\end{tabular}
\caption{Empirical $p$-values for Hyp~\ref{query:uwagowdtomtou}.
The original values of Statistic~\ref{stat:uwagowdtomtou}, with mean and std
over 999 randomized samples are given. The results on randomizations 
$\mathrm{OU}\jointwo \sw(I_1) \jointwo \mathrm{UM} \jointwo \mathrm{MG}$ 
were similar to
$\sw(\mathrm{OU})\jointwo \mathrm{UM} \jointwo \mathrm{MG}$, 
whereas the results on 
$\mathrm{OU}\jointwo \sw(\mathrm{UM}) \jointwo \mathrm{MG}$ 
and $\mathrm{OU}\jointwo \mathrm{UM} \jointwo \sw(\mathrm{MG})$ 
were similar to $\mathrm{OU}\jointwo \mathrm{UM}\jointwo  \sw(I_2)  \jointwo \mathrm{MG}$. Bold
$p$-values are significant with $\sw(\mathrm{OU})$ and nonsignificant with $\sw(I_2)$.}
\label{table:sign:query:uwagowdtomtou}
\end{table}

\begin{query} \label{query:aauwhwmggs}
Average age of the users who have watched movies of a given genre is significant.
\end{query}
\begin{stat} \label{stat:aauwhwmggs}
Weighted average age of the users who have watched movies of the given genre.
\end{stat}
The results of assessing Hyp~\ref{query:aauwhwmggs} are given in Table~\ref{table:sign:query:aauwhwmggs}. 
The empirical $p$-values of the queries depend largely on the type of randomization used. 
By randomizing the ages of the users, that is, $\sw(\mathrm{AU})\jointwo\mathrm{UM}\jointwo\mathrm{MG}$, 
the movies whose average age of 
watchers has originally been around 34 years are not significant. This makes sense when it 
is compared to the average of all users which is 34.1 years. 
Notice that in the query the average is weighted by the number of movies watched by 
the user.  
Thus randomizing the table AU tests the connection between the ages and the users.
Other randomization points tell us that the results on western, 
romance, crime and fantasy are not significant, whereas the results on other genres are significant. 
Thus the inner structure of the User$\times$Movie and Movie$\times$Genre relations 
explain the results of our query. 
The average ages of the users of the genres with a star in Table~\ref{table:sign:query:aauwhwmggs} 
were significant with all types of 
randomizations.

\begin{table}[t]
\small
\centering
\begin{tabular}{l*{5}{@{\hspace{2.8mm}}r}}
\toprule
 & Orig. & \sw(AU) & \sw(UM) & $\sw(I_2)$ & \sw(MG) \\
\midrule
Film-noir* & 35.8 & 0.001 & 0.001 & 0.003 & 0.001 \\
Documentary & 35.0 & \textbf{0.134} & 0.001 & 0.001 & 0.001 \\
Mystery & 34.3 & \textbf{0.197} & 0.001 & 0.004 & 0.001 \\
War & 34.2 & \textbf{0.308} & 0.001 & 0.004 & 0.001 \\
Drama & 34.1 & \textbf{0.493} & 0.001 & 0.001 & 0.001 \\
Western & 33.8 & \textbf{0.307} & 0.001 & \textbf{0.168} & \textbf{0.060} \\
Romance* & 33.4 & 0.024 & 0.001 & 0.039 & 0.002 \\
Musical & 33.0 & 0.016 & \textbf{0.253} & \textbf{0.469} & \textbf{0.257} \\
Crime & 32.6 & 0.001 & 0.001 & \textbf{0.181} & \textbf{0.411} \\
Comedy* & 32.5 & 0.001 & 0.001 & 0.003 & 0.007 \\
Thriller* & 32.2 & 0.001 & 0.001 & 0.003 & 0.004 \\
Adventure* & 32.0 & 0.001 & 0.001 & 0.001 & 0.006 \\
Fantasy & 32.0 & 0.002 & 0.001 & \textbf{0.130} & \textbf{0.164} \\
Children's* & 31.8 & 0.001 & 0.001 & 0.002 & 0.001 \\
Sci-fi* & 31.8 & 0.001 & 0.001 & 0.001 & 0.003 \\
Action* & 31.7 & 0.001 & 0.001 & 0.001 & 0.001 \\
Horror* & 31.1 & 0.001 & 0.001 & 0.001 & 0.001 \\
Animation* & 30.9 & 0.001 & 0.001 & 0.004 & 0.002 \\
\bottomrule
\end{tabular}
\caption{Empirical $p$-values for Hyp~\ref{query:aauwhwmggs}.
The results on randomizations $\mathrm{AU}\jointwo \sw(I_1) \jointwo \mathrm{UM}\jointwo \mathrm{MG}$ 
were similar to $\sw(\mathrm{AU}) \jointwo \mathrm{UM}\jointwo \mathrm{MG}$. Genres with a star are significant with all 
randomizations. Bold $p$-values are non-significant.}
\label{table:sign:query:aauwhwmggs}
\end{table}

\section{Related work}

Obviously, there is a large amount of statistical literature about 
hypothesis testing~\cite{casella01,Good:2000}. 
For the particular case of 
data mining, many papers work on the significance
of association rules and other patterns~\cite{silverstein98beyond,775053}. 
In the recent years, 
the framework of randomizations has been introduced to the data mining community to test
significance of patterns: 
the papers~\cite{cobb03,swaprand-gionisetal} deal with randomizations on binary data, and the work 
in~\cite{conf/sdm/OjalaVKHM08} studies randomizations on real-valued data. 
For another type of approach to measuring $p$-values for patterns, 
see \cite{Webb07}. 
A related work that studies permutations on networks and how this affects
significance of patterns is~\cite{art/KashtanIMA04}.
Sub-sampling methods such as bootstrapping~\cite{Efron79} use randomization to study the properties 
of the underlying distribution instead of testing the data against some null-model. 
Finally, database theory studies mainly query processing and optimization in different
complex data~\cite{DBLP:conf/pods/MoorSAV08,DBLP:conf/pods/JhaRS08}. 
To the best of our knowledge there is no work that directly addresses the problem
presented in this paper.

\section{Conclusions and future work}

We have addressed the problem of assessing the significance of queries
made for the exploratory analysis of relational databases. 
Each query, together with the associated statistic, define the hypothesis to test on our data. 
Our mathematical tool to decide the significance is via randomizations.  
It turns out that in multi-relational data there is no unique way to randomize. 
We propose to randomize tables occurring in the queries one at a time,
and obtain a set of $p$-values for each randomization. 
Each $p$-value tells 
what is the structural impact of the randomized table in the query.
For example, if certain structures or patterns remain after the randomizations,
the answers of a query that rely on such patterns should not be significant. 
Experiments with synthetic data showed that for well defined significant patterns
randomizations uncover which tables from our database are key in significance testing. 
For real datasets, we tested several hypothesis to show the usability of the method.
Still, we found out that in real data 
it is difficult to give a fully satisfactory answer about how to use all the obtained 
$p$-values to conclude the correct inference. Our contribution makes an important 
first step towards
understanding how the structure hidden in the data makes some hypotheses more significant
than others, but still, a lot of interesting future work needs to be done:
study of the combinatorial properties and its connection to the significance
of queries and patterns.

%\bibliographystyle{abbrv}
%\bibliography{./bib}

\end{document}